# Performance Analysis of MIMO-MRC in Double-Correlated Rayleigh Environments

Matthew R. McKay[†,*], Alex J. Grant[‡], and Iain B. Collings[*]

[†]Telecommunications Lab, School of Electrical and Information Engineering, University of Sydney, Australia

[‡]Institute for Telecommunications Research, University of South Australia, Mawson Lakes, Australia

[*]Wireless Technologies Laboratory, ICT Centre, CSIRO, Sydney, Australia

### Abstract

We consider multiple-input multiple-output (MIMO) transmit beamforming systems with maximum ratio combining (MRC) receivers. The operating environment is Rayleigh-fading with *both* transmit and receive spatial correlation. We present exact expressions for the probability density function (p.d.f.) of the output signal-to-noise ratio (SNR), as well as the system outage probability. The results are based on explicit closed-form expressions which we derive for the p.d.f. and c.d.f. of the maximum eigenvalue of double-correlated complex Wishart matrices. For systems with two antennas at either the transmitter or the receiver, we also derive exact closed-form expressions for the symbol error rate (SER). The new expressions are used to prove that MIMO-MRC achieves the maximum available spatial diversity order, and to demonstrate the effect of spatial correlation. The analysis is validated through comparison with Monte-Carlo simulations.

**Corresponding Author:**

Matthew McKay

Sch. of Elec. and Info. Engineering, University of Sydney, NSW 2006, Australia

+61-2-9372 4244 (phone) +61-2-9372 4490 (fax), *mckay@ee.usyd.edu.au*

## I. INTRODUCTION

Multiple-input multiple-output (MIMO) antenna technology can provide significant improvements in capacity [1–4] and error performance [5] over conventional single-antenna technology, without requiring extra power or bandwidth. When channel knowledge is available at both the transmitter and receiver, MIMO transmit beamforming with maximum-ratio combining (MRC) receivers [6] is particularly robust against the severe effects of fading. This robustness is achieved by steering the transmitted signal along the maximum eigenmode of the MIMO channel, resulting in the maximization of the signal-to-noise ratio (SNR) at the MRC output.

Recently, MIMO-MRC has been investigated in uncorrelated and semi-correlated channel scenarios (i.e. where correlation occurs at only one end of the transmission link, or not at all). A key to deriving analytical performance results is to statistically characterize the SNR at the output of the MRC combiner. In [7–11], uncorrelated Rayleigh fading was considered, and the output SNR statistical properties were derived based on maximum eigenvalue statistics of complex central Wishart matrices. In [12], uncorrelated Rician channels were characterized using maximum eigenvalue properties of complex noncentral Wishart matrices. Semi-correlated Rayleigh channels were considered in [13], utilizing properties of semi-correlated Wishart matrices.

In this paper we consider *double-correlated* Rayleigh channels, by first deriving results for the eigenvalue statistics of double-correlated complex Wishart matrices. In practice, double-correlated channels (i.e. with correlation at *both* the transmitter and receiver) commonly occur due to, for example, insufficient scattering around both the transmit and receive terminals, or to closely spaced antennas with respect to the wavelength of the signal. While there are numerous statistical results on general Wishart matrices, there are almost no results for the eigenvalue statistics in the case of double-correlated Wishart matrices. In [14], the joint probability density function (p.d.f.) of the eigenvalues of such matrices was derived in terms of hypergeometric functions of three matrix arguments. In [15], the marginal p.d.f. of an *arbitrary unordered* eigenvalue was derived. Here, we add to these general statistical results, by deriving new exact closed-form expressions for the p.d.f. and cumulative distribution function (c.d.f.) of the *maximum* eigenvalue of double-correlated complex Wishart matrices.

The new general statistical results allow us to consider performance measures for MIMO-MRC in double-correlated Rayleigh environments. In particular, we present explicit expressions for the p.d.f. of the output SNR, and for the outage probability. These expressions are in closed-form, simple, and apply for arbitrary antenna configurations, and over the entire range of SNRs.

We also derive explicit closed-form expressions for the average symbol error rate (SER) with various modulation formats. SER expressions are only currently available in closed-form for uncorrelated Rayleigh channels with $2 \times m$ (i.e. 2 transmit and $m$ receive antennas), or $m \times 2$ systems, where $m \geq 2$ [8, 9]; and all other results either require numerical evaluation of unknown coefficients (e.g. [7]), or evaluation of infinite series. For the SER results in this paper, we also restrict attention to $2 \times m$ and $m \times 2$ systems in order to obtain closed-form solutions, however our results are much more general since they apply for double-correlated channels. The 2-antenna restriction in fact has important practical applications. As discussed in [9], size and cost constraints in realistic cellular MIMO systems will typically limit mobile units to accommodate a maximum of two antennas. Base stations, on the other hand, usually have greater design flexibility, and can easily be implemented with more antennas. As such, $2 \times m$ and $m \times 2$ correspond to uplink and downlink communication scenarios in such systems, respectively.

Even more insights can be gained by analyzing the performance at high SNR. In particular, we formally prove that MIMO-MRC achieves the maximum possible spatial diversity order in double-correlated Rayleigh environments, and show a clear and direct relationship between the SER and the transmit and receive correlation matrices in the high SNR regime.

Finally, we verify the analytical SNR p.d.f., outage probability, and SER expressions by comparing with Monte-Carlo simulations, and examine the impact of correlation in each case. Our results show that the outage probability may increase or decrease as a function of the spatial correlation, whereas the SER tends to vary monotonically.

## II. STATISTICAL PROPERTIES OF DOUBLE-CORRELATED COMPLEX WISHART MATRICES

This section presents new general statistical properties of complex double-correlated Wishart matrices, which will be used in the subsequent performance analysis of MIMO-MRC.

## A. Cumulative Distribution Function (C.D.F.)

The following theorem presents the c.d.f. of the maximum eigenvalue of double-correlated complex Wishart matrices. This will be used for deriving the outage probability of MIMO-MRC in double-correlated Rayleigh channels.

*Theorem 1:* Let $\mathbf{X} \sim \mathcal{CN}_{m,n}\left(\mathbf{0}_{m \times n}, \mathbf{\Sigma} \otimes \mathbf{\Omega}\right)$, where $n \leq m$, and $\mathbf{\Omega} \in \mathcal{C}^{n \times n}$ and $\mathbf{\Sigma} \in \mathcal{C}^{m \times m}$ are Hermitian positive-definite matrices with eigenvalues $\omega_1 < \ldots < \omega_n$ and $\sigma_1 < \ldots < \sigma_m$ respectively. Then the c.d.f. of the maximum eigenvalue $\lambda_m$ of the double-correlated complex Wishart matrix $\mathbf{X}^\dagger \mathbf{X}$ is given by

$$F_{\lambda_m}(x) = \frac{(-1)^n \Gamma_n(n) \det(\mathbf{\Omega})^{n-1} \det(\mathbf{\Sigma})^{m-1} \det(\mathbf{\Psi}(x))}{\Delta_n(\mathbf{\Omega}) \Delta_m(\mathbf{\Sigma})(-x)^{n(n-1)/2}} \tag{1}$$

where $\Gamma_n(\cdot)$ is the normalized complex multivariate gamma function, defined as[1]

$$\Gamma_n(n) = \prod_{i=1}^{n} \Gamma(n - i + 1) \tag{2}$$

and $\Delta_m(\cdot)$ is a Vandermonde determinant in the eigenvalues of the $m$-dimensional matrix argument[2], given by

$$\Delta_m(\mathbf{\Sigma}) = \prod_{i<j}^{m} (\sigma_j - \sigma_i) . \tag{3}$$

Also, $\mathbf{\Psi}(x)$ is an $m \times m$ matrix with $(i,j)^{\text{th}}$ element

$$(\mathbf{\Psi}(x))_{i,j} = \begin{cases} \left(\frac{1}{\sigma_j}\right)^{m-i} & \text{for } i \leq \tau \\ e^{-\frac{x}{\omega_{i-\tau}\sigma_j}} \mathcal{P}\left(m; -\frac{x}{\omega_{i-\tau}\sigma_j}\right) & \text{for } i > \tau \end{cases} \tag{4}$$

where $\tau = m - n$, and

$$\mathcal{P}\left(\ell; y\right) = 1 - e^{-y} \sum_{k=0}^{\ell-1} y^k / k! \tag{5}$$

is the regularized lower incomplete gamma function.

*Proof:* See the Appendix.                                         ∎

---

[1] Note that this is related to the standard complex multivariate gamma function $\tilde{\Gamma}_n(n)$ (as defined in [16]) via $\Gamma_n(n) = \pi^{-n(n-1)/2} \tilde{\Gamma}_n(n)$.

[2] For notational convenience, we will sometimes give a set in place of a matrix in the argument of $\Delta_m(\cdot)$, as a shorthand. Specifically, for the set $\mathcal{A}$, we use $\Delta_m(\mathcal{A})$ as shorthand for $\Delta_m(\text{diag}(\mathcal{A}))$.

*Corollary 1:* For the case $n = 2, m \geq 2$, (1) reduces to

$$F_{\lambda_m}(x) = \frac{\det(\boldsymbol{\Omega})}{\Delta_2(\boldsymbol{\Omega})\,\Delta_m(\boldsymbol{\Sigma})} \sum_{p=1}^{m} \sum_{t=1, t \neq p}^{m} (-1)^{p+\phi(t)} (\sigma_p \sigma_t)^{m-1} \Delta_{m-2}\left(\sigma^{[p,t]}\right) Q_{p,t}(x) \qquad (6)$$

where

$$\phi(t) = \begin{cases} t & , \ t < p \\ t-1 & , \ t > p \end{cases} \qquad (7)$$

and $\sigma^{[p,t]} = \{\sigma_i ; i \in \{1, \ldots, m\} \setminus \{p, t\}\}$, and

$$Q_{p,t}(x) = \frac{1}{x} e^{-\frac{x}{\omega_2 \sigma_p}} \mathcal{P}\left(m; -\frac{x}{\omega_2 \sigma_p}\right) e^{-\frac{x}{\omega_1 \sigma_t}} \mathcal{P}\left(m; -\frac{x}{\omega_1 \sigma_t}\right). \qquad (8)$$

*Proof:* The proof follows by applying Laplace's Expansion to the last two rows of $\det(\boldsymbol{\Psi}(x))$ in (1), and then simplifying the resulting (Vandermonde-type) minors using (56). ∎

*Corollary 2:* For the case $n = m = 2$, (1) reduces to the simple expression

$$F_{\lambda_m}(x) = \frac{\omega_1 \omega_2 \sigma_1 \sigma_2}{x(\sigma_2 - \sigma_1)(\omega_2 - \omega_1)} \sum_{i=1}^{2} (-1)^i \prod_{j=1}^{2} \left( e^{-\frac{x}{\omega_{|i-j|+1}\sigma_j}} + \frac{x}{\omega_{|i-j|+1}\sigma_j} - 1 \right). \qquad (9)$$

*Proof:* The proof is straightforward and is omitted. ∎

## B. Probability Density Function (P.D.F.)

The following theorem presents the p.d.f. of the maximum eigenvalue of double-correlated complex Wishart matrices. This will be used for deriving the p.d.f. of the output SNR of MIMO-MRC in double-correlated Rayleigh channels.

*Theorem 2:* Let $\mathbf{X} \sim \mathcal{CN}_{m,n}(\mathbf{0}_{m \times n}, \boldsymbol{\Sigma} \otimes \boldsymbol{\Omega})$, where $n \leq m$, and $\boldsymbol{\Omega} \in \mathcal{C}^{n \times n}$ and $\boldsymbol{\Sigma} \in \mathcal{C}^{m \times m}$ are Hermitian positive-definite matrices with eigenvalues $\omega_1 < \ldots < \omega_n$ and $\sigma_1 < \ldots < \sigma_m$ respectively. Then the p.d.f. of the maximum eigenvalue $\lambda_m$ of the double-correlated complex Wishart matrix $\mathbf{X}^\dagger \mathbf{X}$ is given by

$$f_{\lambda_m}(\lambda_m) = \frac{(-1)^{n+1} \Gamma_n(n) \det(\boldsymbol{\Omega})^{n-1} \det(\boldsymbol{\Sigma})^{m-1}}{\Delta_n(\boldsymbol{\Omega})\,\Delta_m(\boldsymbol{\Sigma})\,(-\lambda_m)^{n(n-1)/2}}$$
$$\times \left( \frac{n(n-1)\det(\boldsymbol{\Psi}(\lambda_m))}{2\lambda_m} + \sum_{\ell=\tau+1}^{m} \det(\boldsymbol{\Psi}_\ell(\lambda_m)) \right) \qquad (10)$$

where $\boldsymbol{\Psi}_\ell(\lambda_m)$ is an $m \times m$ matrix with $(i,j)^{\text{th}}$ element

$$(\mathbf{\Psi}_\ell(\lambda_m))_{i,j} = \begin{cases} (\mathbf{\Psi}(\lambda_m))_{i,j} & \text{for } i \neq \ell \\ \dfrac{e^{-\frac{\lambda_m}{\omega_{i-\tau}\sigma_j}}}{\omega_{i-\tau}\sigma_j}\mathcal{P}\left(m-1; \dfrac{-\lambda_m}{\omega_{i-\tau}\sigma_j}\right) & \text{for } i = \ell \end{cases} \tag{11}$$

and where $(\mathbf{\Psi}(\lambda_m))_{i,j}$ is defined in (4).

*Proof:* The result follows by differentiating (1) with respect to $x$.  ∎

## III. MIMO-MRC System Model

Consider $N_t$ transmit and $N_r$ receive antennas, where the $N_r \times 1$ received vector is

$$\mathbf{r} = \sqrt{\bar{\gamma}}\mathbf{H}\mathbf{w}x + \mathbf{n} \tag{12}$$

where $x$ is the transmitted symbol with $E\left[|x|^2\right] = 1$, $\mathbf{w}$ is the beamforming vector (specified below) with $E\left[\|\mathbf{w}\|^2\right] = 1$, $\mathbf{n}$ is noise $\sim \mathcal{CN}_{N_r,1}\left(\mathbf{0}_{N_r \times 1}, \mathbf{I}_{N_r}\right)$, and $\bar{\gamma}$ is the transmit SNR. Also, $\mathbf{H}$ is the $N_r \times N_t$ channel matrix, assumed to be flat spatially-correlated Rayleigh fading, and is decomposed according to the common *kronecker* structure (as in [4, 14, 17, 18], among others) as

$$\mathbf{H} = \mathbf{R}^{\frac{1}{2}}\mathbf{H}_w\mathbf{S}^{\frac{1}{2}} \quad \sim \ \mathcal{CN}_{N_r,N_t}\left(\mathbf{0}_{N_r \times N_t}, \mathbf{R} \otimes \mathbf{S}\right) \tag{13}$$

where $\mathbf{R}$ and $\mathbf{S}$ are the receive and transmit correlation matrices respectively, with unit diagonal entries, and $\mathbf{H}_w \sim \mathcal{CN}_{N_r,N_t}\left(\mathbf{0}_{N_r \times N_t}, \mathbf{I}_{N_r} \otimes \mathbf{I}_{N_t}\right)$.

The receiver employs the principle of MRC to give

$$\mathbf{z} = \mathbf{w}^\dagger\mathbf{H}^\dagger\mathbf{r} = \sqrt{\bar{\gamma}}\mathbf{w}^\dagger\mathbf{H}^\dagger\mathbf{H}\mathbf{w}x + \mathbf{w}^\dagger\mathbf{H}^\dagger\mathbf{n} \quad . \tag{14}$$

Therefore, the SNR at the output of the combiner is easily derived as

$$\gamma = \bar{\gamma}\mathbf{w}^\dagger\mathbf{H}^\dagger\mathbf{H}\mathbf{w} \quad . \tag{15}$$

The BF vector $\mathbf{w}$ is chosen to maximize this instantaneous output SNR, thereby minimizing the error probability. It is well known that the optimum BF vector $\mathbf{w}_{\text{opt}}$ is the eigenvector corresponding to the maximum eigenvalue $\lambda_m$ of $\mathbf{H}^\dagger\mathbf{H}$. In this case, the output SNR (15) becomes

$$\gamma = \bar{\gamma}\mathbf{w}_{\text{opt}}^\dagger\mathbf{H}^\dagger\mathbf{H}\mathbf{w}_{\text{opt}} = \bar{\gamma}\lambda_m \quad . \tag{16}$$

Clearly the output SNR (and therefore the performance) of MIMO-MRC depends directly on the statistical properties of $\lambda_m$.

In the previous section we presented results which apply directly to $\lambda_m$ for the case when $N_r \geq N_t$ (i.e. $\mathbf{H}^\dagger \mathbf{H}$ is full-rank), by setting $\mathbf{\Omega} = \mathbf{S}$, $\mathbf{\Sigma} = \mathbf{R}$, $n = N_t$, and $m = N_r$. The results of the previous section also apply to the case when $N_r < N_t$, (i.e. $\mathbf{H}\mathbf{H}^\dagger$ is full-rank), by setting $\mathbf{\Omega} = \mathbf{R}$, $\mathbf{\Sigma} = \mathbf{S}$, $n = N_r$, and $m = N_t$; since the maximum eigenvalue is the same in both cases.

## IV. Performance Analysis of MIMO-MRC in Double-Correlated Channels

### A. Statistical Characterization of the Output SNR

We now characterize the statistics of the output SNR $\gamma$. Using (16), and making a simple change of variables to (10) we obtain the p.d.f. of $\gamma$, given in closed-form by

$$f_\gamma(\gamma) = \frac{(-1)^n \Gamma_n(n) \det(\mathbf{\Omega})^{n-1} \det(\mathbf{\Sigma})^{m-1}}{\Delta_n(\mathbf{\Omega}) \, \Delta_m(\mathbf{\Sigma})} \left( -\frac{\bar{\gamma}}{\gamma} \right)^{n(n-1)/2}$$
$$\times \left( \frac{n(1-n) \det\left( \mathbf{\Psi}\left( \frac{\gamma}{\bar{\gamma}} \right) \right) \bar{\gamma}}{2\gamma} + \sum_{\ell=\tau+1}^{m} \det\left( \mathbf{\Psi}_\ell\left( \frac{\gamma}{\bar{\gamma}} \right) \right) \right) . \quad (17)$$

The outage probability is an important quality of service measure, defined as the probability that $\gamma$ drops below an acceptable SNR threshold $\gamma_{\text{th}}$. It is obtained from (1) and (15) as follows

$$F_\gamma(\gamma_{\text{th}}) = \Pr(\gamma \leq \gamma_{\text{th}}) = F_{\lambda_m}\left( \frac{\gamma_{\text{th}}}{\bar{\gamma}} \right)$$
$$= \frac{(-1)^n \Gamma_n(n) \det(\mathbf{\Omega})^{n-1} \det(\mathbf{\Sigma})^{m-1} \det\left( \mathbf{\Psi}\left( \frac{\gamma_{\text{th}}}{\bar{\gamma}} \right) \right)}{\Delta_n(\mathbf{\Omega}) \, \Delta_m(\mathbf{\Sigma}) \left( -\frac{\gamma_{\text{th}}}{\bar{\gamma}} \right)^{n(n-1)/2}} . \quad (18)$$

In Section V we show that the outage probability can increase or decrease with the spatial correlation, depending on the value of $\gamma_{\text{th}}$.

### B. Symbol Error Rate Analysis

In this subsection we derive closed-form expressions for the average symbol error rate (SER) of MIMO-MRC in double-correlated Rayleigh channels, with various modulation formats. As discussed in Section I, we consider the practical special cases $2 \times m$ and $m \times 2$. Our results apply for all general modulation formats that have a SER expression of the form

$$P_s = E_\gamma \left[ aQ \left( \sqrt{2b\gamma} \right) \right] \tag{19}$$

where $Q(\cdot)$ is the Gaussian Q-function, and $a$ and $b$ are modulation-specific constants. Such modulation formats include BPSK ($a = 1$, $b = 1$); BFSK with orthogonal signalling ($a = 1$, $b = 0.5$) or minimum correlation ($a = 1$, $b = 0.715$); and $M$−ary PAM ($a = 2(M-1)/M$, $b = 3/(M^2 - 1)$). Our results also provide the approximate SER for those other formats for which (19) is an approximation, e.g. $M$-ary PSK ($a = 2, b = \sin^2(\pi/M)$) [19, Eq. 5.2-61].

A common useful approach for evaluating SERs of the form (19), is to first evaluate the moment generating function (m.g.f.) of $\gamma$, and then apply the well-known m.g.f.-SER relationships given in [20]. In the context of MIMO-MRC, this approach was used in [7, 9] to evaluate SERs in uncorrelated Rayleigh channels. Although it is possible to evaluate the output SNR m.g.f. for double-correlated channels considered in this paper, unfortunately the resulting expression has significant convergence problems.

As such, we adopt an alternative approach. We begin by generalizing a result presented in [13], which directly related the SER of MIMO-MRC in semi-correlated channels with BPSK to the c.d.f. of the output SNR. We start by noting that the Gaussian Q-function can be written as

$$Q(x) = \frac{1}{\sqrt{\pi}} \int_{x/\sqrt{2}}^{\infty} e^{-v^2/2} \, \mathrm{d}v \tag{20}$$

and substitute into (19) to give

$$P_s = \frac{a}{\sqrt{\pi}} \int_0^\infty \left[ \int_{\sqrt{bu}}^\infty e^{-v^2} \mathrm{d}v \right] f_\gamma(u) \, \mathrm{d}u \ . \tag{21}$$

Now we apply (definite) integration by parts to (21). To do this we integrate $f_\gamma(u)$, and differentiate the quantity in square brackets as follows

$$\begin{aligned}
\frac{\mathrm{d}}{\mathrm{d}u} \left[ \int_{\sqrt{bu}}^\infty e^{-v^2} \mathrm{d}v \right] &= -\frac{\mathrm{d}}{\mathrm{d}u} \left[ \int_0^{\sqrt{bu}} e^{-v^2} \mathrm{d}v \right] \\
&= -\frac{\mathrm{d}}{\mathrm{d}u} \sqrt{bu} \left( \frac{\mathrm{d}}{\mathrm{d}x} \left[ \int_0^x e^{-v^2} \mathrm{d}v \right] \right)_{x=\sqrt{bu}} \\
&= -\frac{1}{2} \sqrt{\frac{b}{u}} e^{-bu} \tag{22}
\end{aligned}$$

where the first line followed by noting that

$$\int_0^\infty e^{-v^2} \mathrm{d}v = \frac{\sqrt{\pi}}{2} \tag{23}$$

and the second line followed from the Chain Rule, and last line was obtained using the Second Fundamental Theorem of Calculus. This gives

$$P_s = \frac{a\sqrt{b}}{2\sqrt{\pi}} \int_0^\infty \frac{e^{-bu}}{\sqrt{u}} F_\gamma(u) \, \mathrm{d}u. \tag{24}$$

Note that (24) is a generalization of [13, Eq. 32] to SER expressions of the form (19).

Now we can directly substitute (18) into (24) and integrate term by term. This procedure, however, is problematic since the term-by-term integrals are divergent (although the overall result converges). To circumvent this problem, we first write (24) in terms of the c.d.f. of $\lambda_m$, given in Corollary 1, as follows

$$P_s = \frac{a\sqrt{b}}{2\sqrt{\pi}} \int_0^\infty \frac{e^{-bu}}{\sqrt{u}} F_{\lambda_m}\left(\frac{u}{\bar{\gamma}}\right) \mathrm{d}u \tag{25}$$

and derive an alternative expression for $Q_{p,t}(x)$ in (8).

Expanding the exponentials into power series, and using (5), we manipulate (8) as follows

$$\begin{aligned}
Q_{p,t}(x) &= \frac{1}{x} \left( \sum_{k=m}^\infty \frac{\left(-\frac{x}{\omega_2 \sigma_p}\right)^k}{k!} \right) \left( \sum_{k=m}^\infty \frac{\left(-\frac{x}{\omega_1 \sigma_t}\right)^k}{k!} \right) \\
&= \frac{1}{x} \sum_{k_1=m}^\infty \sum_{k_2=m}^\infty \frac{(-x)^{k_1+k_2} \left(\frac{1}{\omega_2 \sigma_p}\right)^{k_1} \left(\frac{1}{\omega_1 \sigma_t}\right)^{k_2}}{k_1! k_2!} \\
&= -\sum_{k=2m}^\infty (-x)^{k-1} \mathcal{S}_k
\end{aligned} \tag{26}$$

where

$$\mathcal{S}_k = \sum_{\ell=m}^{k-m} \frac{\left(\frac{1}{\omega_2 \sigma_p}\right)^\ell \left(\frac{1}{\omega_1 \sigma_t}\right)^{k-\ell}}{\ell!(k-\ell)!} \ . \tag{27}$$

Now, substituting (26) into (6) and (25), the SER can be written as

$$P_s = \frac{a\sqrt{b}}{2\sqrt{\pi}} \frac{\det(\mathbf{\Omega})}{\Delta_2(\mathbf{\Omega}) \Delta_m(\mathbf{\Sigma})} \sum_{p=1}^m \sum_{t=1, t \neq p}^m (-1)^{p+\phi(t)} (\sigma_p \sigma_t)^{m-1} \Delta_{m-2}\left(\sigma^{[p,t]}\right) \mathcal{I} \tag{28}$$

where

$$
\begin{aligned}
\mathcal{I} &= \int_0^\infty \frac{e^{-bu}}{\sqrt{u}} Q_{p,t}\left(\frac{u}{\bar{\gamma}}\right) \mathrm{d}u \\
&= -\sum_{k=2m}^\infty \mathcal{S}_k \left(-\frac{1}{\bar{\gamma}}\right)^{k-1} \int_0^\infty u^{k-3/2} e^{-bu} \mathrm{d}u \\
&= \bar{\gamma}\sqrt{b} \sum_{k=2m}^\infty \mathcal{S}_k \left(-\frac{1}{b\bar{\gamma}}\right)^k \Gamma\left(k-1/2\right)
\end{aligned}
\tag{29}
$$

where the last line followed by applying the integration identity [21, Eq. 3.381.4]. To remove the infinite summation from (29), we apply the Binomial Theorem to express $\mathcal{S}_k$ in (27) as

$$
\mathcal{S}_k = \frac{1}{k!}\left(\left(\frac{1}{\omega_2\sigma_p} + \frac{1}{\omega_1\sigma_t}\right)^k - \sum_{\ell=0}^{m-1}\binom{k}{\ell}\left(\frac{1}{\omega_2\sigma_p}\right)^\ell\left(\frac{1}{\omega_1\sigma_t}\right)^{k-\ell} \right.
$$
$$
\left. - \sum_{\ell=k-m+1}^{k}\binom{k}{\ell}\left(\frac{1}{\omega_2\sigma_p}\right)^\ell\left(\frac{1}{\omega_1\sigma_t}\right)^{k-\ell}\right).
\tag{30}
$$

Now, performing the change of variables $\ell \to (k-\ell)$ in the last summation in (30), and using the property $\binom{k}{\ell} = \binom{k}{k-\ell}$, we can write

$$
\mathcal{S}_k = \frac{1}{k!}\left(\left(\frac{1}{\omega_2\sigma_p} + \frac{1}{\omega_1\sigma_t}\right)^k - \sum_{\ell=0}^{m-1}\binom{k}{\ell}\left(\left(\frac{1}{\omega_2\sigma_p}\right)^\ell\left(\frac{1}{\omega_1\sigma_t}\right)^{k-\ell} + \left(\frac{1}{\omega_1\sigma_t}\right)^\ell\left(\frac{1}{\omega_2\sigma_p}\right)^{k-\ell}\right)\right).
\tag{31}
$$

Note that the key advantage of writing $\mathcal{S}_k$ in this way, as opposed to (27), is that we have removed the dependence of the summation limits on $k$. We now substitute (31) into (29) and simplify to obtain

$$
\mathcal{I} = \bar{\gamma}\sqrt{b}\left(\eta\left(0, -\frac{1}{\bar{\gamma}b}\left(\frac{1}{\omega_2\sigma_p} + \frac{1}{\omega_1\sigma_t}\right)\right) - \sum_{\ell=0}^{m-1}\frac{1}{\ell!}\frac{\eta\left(\ell, -\frac{1}{\bar{\gamma}b\omega_2\sigma_p}\right)}{\left(\bar{\gamma}b\omega_1\sigma_t\right)^\ell} - \sum_{\ell=0}^{m-1}\frac{1}{\ell!}\frac{\eta\left(\ell, -\frac{1}{\bar{\gamma}b\omega_1\sigma_t}\right)}{\left(\bar{\gamma}b\omega_2\sigma_p\right)^\ell}\right)
\tag{32}
$$

where

$$
\begin{aligned}
\eta(\ell, y) &= \sum_{k=2m}^\infty \frac{y^{k-\ell}\,\Gamma\left(k-1/2\right)}{(k-\ell)!} \\
&= \sum_{k=2m-\ell}^\infty \frac{y^k\,\Gamma\left(k+\ell-1/2\right)}{k!} \\
&= \Gamma\left(\ell-1/2\right){}_1F_0\left(\ell-1/2;y\right) - \sum_{k=0}^{2m-\ell-1}\frac{y^k\,\Gamma\left(k+\ell-1/2\right)}{k!}
\end{aligned}
\tag{33}
$$

where $_1F_0(\cdot)$ is the binomial hypergeometric function. Noting that $_1F_0(a;y) = {}_2F_1(a;b;b;y)$, and using [22, Eq. 15.1.8], we obtain

$$\eta(\ell, y) = \Gamma(\ell - 1/2)(1-y)^{1/2-\ell} - \sum_{k=0}^{2m-\ell-1} \frac{y^k \, \Gamma(k+\ell-1/2)}{k!} \; . \tag{34}$$

Now we apply the identity

$$\Gamma(k+1/2) = \frac{(2k-1)!! \sqrt{\pi}}{2^k} \tag{35}$$

where

$$(2k-1)!! \stackrel{\triangle}{=} 1 \times 3 \times \ldots \times (2k-1) \tag{36}$$

for $k > 0$, and define $(-1)!! = 1$ and $(-3)!! = -1$, which yields

$$\eta(\ell, y) = \frac{\sqrt{\pi}}{2^{\ell-1}} \tilde{\eta}(\ell, y) \tag{37}$$

where

$$\tilde{\eta}(\ell, y) = (2\ell - 3)!! (1-y)^{1/2-\ell} - \sum_{k=0}^{2m-\ell-1} \left(\frac{y}{2}\right)^k \frac{(2(k+\ell)-3)!!}{k!} \; . \tag{38}$$

Finally, substituting (37) into (32) and (28) yields the desired closed-form SER expression

$$P_s = \frac{\det(\mathbf{\Omega}) \, ab\bar{\gamma}}{\Delta_2(\mathbf{\Omega}) \, \Delta_m(\mathbf{\Sigma})} \sum_{p=1}^{m} \sum_{t=1, t\neq p}^{m} (-1)^{p+\phi(t)} (\sigma_p \sigma_t)^{m-1} \, \Delta_{m-2}(\sigma^{[p,t]})$$
$$\times \left( \tilde{\eta}\left(0, -\frac{1}{\bar{\gamma}b}\left(\frac{1}{\omega_2\sigma_p} + \frac{1}{\omega_1\sigma_t}\right)\right) - \sum_{\ell=0}^{m-1} \frac{1}{\ell!} \frac{\tilde{\eta}\left(\ell, -\frac{1}{\bar{\gamma}b\omega_2\sigma_p}\right)}{(2\bar{\gamma}b\omega_1\sigma_t)^\ell} - \sum_{\ell=0}^{m-1} \frac{1}{\ell!} \frac{\tilde{\eta}\left(\ell, -\frac{1}{\bar{\gamma}b\omega_1\sigma_t}\right)}{(2\bar{\gamma}b\omega_2\sigma_p)^\ell} \right) \; . \tag{39}$$

Note that (39) is a simple finite closed-form expression (involving only polynomial terms in $\bar{\gamma}$), which can be evaluated easily and efficiently.

### C. High SNR SER Analysis

We now analyze the SER performance in the high SNR regime in order to derive the diversity order of the system.

Consider the SER expression given by (28) and (29). It can easily be shown that, irrespective of the SNR, the first term in (29) (i.e. $k = 2m$) cancels with terms in (28) to give zero contribution to $P_s$. Therefore the summation in (29) could equally well be written starting from $k = 2m+1$. Since as $\bar{\gamma} \to \infty$, the infinite series in (29) is dominated by the low order terms, in the high

SNR regime we need only to consider $k = 2m + 1$. To proceed, we take the $k = 2m + 1$ term in the series and use (27) to write (29) at high SNR as follows

$$\mathcal{I}^{\infty} = -\frac{(\bar{\gamma}b)^{-2m}\,\Gamma(2m+1/2)}{\sqrt{b}\,m!(m+1)!}\left(\left(\frac{1}{\omega_2\sigma_p}\right)^m\left(\frac{1}{\omega_1\sigma_t}\right)^{m+1} + \left(\frac{1}{\omega_2\sigma_p}\right)^{m+1}\left(\frac{1}{\omega_1\sigma_t}\right)^m\right)$$

$$= -\frac{(\bar{\gamma}b)^{-2m}\,\Gamma(2m+1/2)}{\sqrt{b}\,m!(m+1)!}\frac{1}{\det\left(\boldsymbol{\Omega}\right)^m(\sigma_p\sigma_t)^m}\left(\frac{1}{\omega_1\sigma_t} + \frac{1}{\omega_2\sigma_p}\right)\ . \tag{40}$$

Substituting (40) into (28) gives $P_s$ at high SNR as follows

$$P_s^{\infty} = -\frac{a\,\Gamma(2m+1/2)\,(\bar{\gamma}b)^{-2m}}{2\sqrt{\pi}\,m!(m+1)!\Delta_2\left(\boldsymbol{\Omega}\right)\Delta_m\left(\boldsymbol{\Sigma}\right)\det\left(\boldsymbol{\Omega}\right)^{m-1}}\tilde{\mathcal{S}} \tag{41}$$

where

$$\tilde{S} = \sum_{p=1}^{m}\sum_{t=1,t\neq p}^{m}(-1)^{p+\phi(t)}\frac{\Delta_{m-2}\left(\sigma^{[p,t]}\right)}{\sigma_p\sigma_t}\left(\frac{1}{\omega_1\sigma_t} + \frac{1}{\omega_2\sigma_p}\right)\ . \tag{42}$$

Now, it can be shown that

$$\sum_{p=1}^{m}\sum_{t=1,t\neq p}^{m}(-1)^{p+\phi(t)}\frac{\Delta_{m-2}\left(\sigma^{[p,t]}\right)}{\sigma_p\sigma_t^2}\ \text{ and }\ \sum_{p=1}^{m}\sum_{t=1,t\neq p}^{m}(-1)^{p+\phi(t)}\frac{\Delta_{m-2}\left(\sigma^{[p,t]}\right)}{\sigma_p^2\sigma_t}$$

are Laplace expansions of $(-1)^{m-2}\Delta_m(\boldsymbol{\Sigma})$ and $(-1)^{m-1}\Delta_m(\boldsymbol{\Sigma})$ respectively. As such, (42) can be written as

$$\tilde{S} = -\left(\frac{1}{\omega_1} - \frac{1}{\omega_2}\right)\frac{\Delta_m(\boldsymbol{\Sigma})}{\det\left(\boldsymbol{\Sigma}\right)^2}$$

$$= -\frac{\Delta_2(\boldsymbol{\Omega})\Delta_m(\boldsymbol{\Sigma})}{\det\left(\boldsymbol{\Omega}\right)\det\left(\boldsymbol{\Sigma}\right)^2} \tag{43}$$

where the second line followed from (56). Finally, substituting (43) into (41) and simplifying using (35), we obtain the desired high SNR SER result

$$P_s^{\infty} = \frac{a\,(4m-1)!!}{b^{2m}\,2^{2m+1}\,m!(m+1)!\det\left(\boldsymbol{\Omega}\right)^m\det\left(\boldsymbol{\Sigma}\right)^2}\,\bar{\gamma}^{-2m}\ . \tag{44}$$

Therefore, clearly MIMO-MRC achieves the maximum possible spatial diversity order of $2m$ in double-correlated Rayleigh channels. Also, using Hadamard's inequality [23, Th. 16.8.2], and the fact that the diagonal elements of $\boldsymbol{\Omega}$ and $\boldsymbol{\Sigma}$ are unity, it can be easily shown that

$$0 \leq \det\left(\boldsymbol{\Omega}\right) \leq 1\ \text{ and }\ 0 \leq \det\left(\boldsymbol{\Sigma}\right) \leq 1 \tag{45}$$

with equality in the upper limit only when the correlation matrices are identity matrices. This proves that the presence of spatial correlation (at either end) yields a net reduction in error performance in the high SNR regime.

## V. Numerical Results

While the analytic results in this paper apply to arbitrary channel correlation matrices, for our numerical studies we construct the correlation matrices using the practical channel model presented in [24]. The model assumes that there are uniform linear arrays at both the transmitter and receiver. Let us denote the relative antenna spacing between adjacent antennas (measured in number of wavelengths) as $d_r$ at the receiver and $d_t$ at the transmitter. Also, define $\theta_r, \theta_t, \sigma_r^2$ and $\sigma_t^2$ as the mean angle of arrival (AoA), mean angle of departure (AoD), receive angle spread and transmit angle spread respectively, and let $\underline{\theta}_r = \theta_r + \hat{\theta}_r$ and $\underline{\theta}_t = \theta_t + \hat{\theta}_t$ denote the actual AoA and AoD, with $\hat{\theta}_r \sim \mathcal{N}\left(0, \sigma_r^2\right)$ and $\hat{\theta}_t \sim \mathcal{N}\left(0, \sigma_t^2\right)$. With these definitions, the $(p, q)^{\text{th}}$ entry of $\mathbf{R}$ and $\mathbf{S}$ is given by

$$\mathbf{R}_{p,q} = e^{-j2\pi(q-p)d_r \cos(\theta_r)} e^{-\frac{1}{2}(2\pi(q-p)d_r \sin(\theta_r)\sigma_r)^2}, \quad \mathbf{S}_{p,q} = e^{-j2\pi(p-q)d_t \cos(\theta_t)} e^{-\frac{1}{2}(2\pi(p-q)d_t \sin(\theta_t)\sigma_t)^2}.$$

Recall that $\mathbf{R}$ and $\mathbf{S}$ are directly related to $\Omega$ and $\Sigma$ as discussed at the end of Section III. For all results in this section we assume $d_r = d_t = \frac{1}{2}$ and $\theta_r = \theta_t = \frac{\pi}{2}$.

Fig. 1 shows the p.d.f. of the output SNR of MIMO-MRC with various antenna configurations. The 'Analytical' curves are from (17), and clearly agree with the Monte-Carlo simulated p.d.f.s. Moreover, we observe that both the mean and variance of the output SNR increase with the number of antennas.

Fig. 2 shows the (analytical) p.d.f. of the output SNR, comparing various correlation scenarios. We see that for $4 \times 4$ antennas, extra correlation increases the spread of the SNR around the mean. This effect was also observed for semi-correlated channels in [13]. For the $2 \times 2$ case, the correlation has less effect.

Fig. 3 shows the outage probability of MIMO-MRC, with the same antenna configurations as in Fig. 1. The 'Analytical' curves are from (18), and agree precisely with Monte-Carlo simulated curves. We see that the outage probability is significantly improved as the number of antennas are increased.

Fig. 4 shows (analytical) outage probability curves, comparing different correlation scenarios.

We see that for both antenna configurations, the correlation increases the outage probability at low SNR thresholds, and decreases the outage probability (thereby improving system performance) at high SNR thresholds. These general trends were also previously observed for semi-correlated channels in [13]. Moreover, we see that the cross-over point of the different correlated curves occurs at lower outage probabilities as the numbers of antennas increase.

Fig. 5 shows the SER of MIMO-MRC with BPSK modulation, for various antenna configurations. The 'Analytical' curves are from (39) with $a = 1$ and $b = 1$, and match exactly with the Monte-Carlo simulated SERs. The 'Analytical (High SNR)' curves are from (44). Clearly they converge to the exact SER in the high SNR regime, confirming that the maximum possible diversity order is achieved.

Fig. 6 shows the SER of MIMO-MRC with 4-PAM modulation, for various antenna configurations. The 'Analytical' curves were generated from (39) with $a = 1.5$ and $b = 0.2$. Again we see an exact agreement with the Monte-Carlo simulated curves in all cases.

Fig. 7 shows the SER of MIMO-MRC with QPSK modulation, for various antenna configurations. The 'Analytical' curves were generated based on (39) with $a = 2$ and $b = 0.5$. As discussed in Section IV-B, (39) only provides an approximation for QPSK, however we see that this approximation is accurate for all but very low SNRs. In particular, for all SERs of practical interest (i.e. $P_s < 0.01$), the analytical curves match almost exactly with the simulated curves.

Fig. 8 uses the analytical expressions to examine the effect of correlation on the SER. Results are presented for BPSK modulation. We see that for both antenna configurations, the SER increases monotonically with the level of correlation. This agrees with the high SNR predictions from (44) and (45).

## VI. Conclusion

We have examined the performance of MIMO-MRC in double-correlated Rayleigh fading environments. Our results are based on exact closed-form expressions which we derived for the p.d.f. and c.d.f. of the maximum eigenvalue of double-correlated complex Wishart matrices. We showed that the outage performance may increase or decrease due to the presence of spatial correlation, depending on the average SNR. When the minimum number of transmit and receive antennas is two, we proved that MIMO-MRC achieves the maximum available spatial diversity order in double-correlated channels.

<div align="center">APPENDIX</div>

### A. Proof of Theorem 1

First consider the $n = m$ case of square random matrices $\mathbf{X} \sim \mathcal{CN}_{m,m}(\mathbf{0}_{m \times m}, \mathbf{\Sigma} \otimes \mathbf{\Omega})$, and let $\lambda_1 < \ldots < \lambda_m$ be the non-zero eigenvalues of $\mathbf{X}^\dagger \mathbf{X}$. The c.d.f. of $\lambda_m$ is obtained using

$$F_{\lambda_m}(x) = \int_{\mathcal{D}} f(\mathbf{\Lambda}) \, d\mathbf{\Lambda} \tag{46}$$

where $\mathbf{\Lambda} = \mathrm{diag}\{\lambda_1, \ldots, \lambda_m\}$, $f(\mathbf{\Lambda})$ is the joint p.d.f. of $\lambda_1, \ldots, \lambda_m$, and $\mathcal{D} = \{0 \leq \lambda_1 \leq \ldots \leq \lambda_m < x\}$. It was shown in [14] that

$$f(\mathbf{\Lambda}) = \frac{{}_0\tilde{F}_0\left(-\mathbf{\Omega}^{-1}, \mathbf{\Sigma}^{-1}, \mathbf{\Lambda}\right) \Delta_m(\mathbf{\Lambda})^2}{\Gamma_m(m)^2 \, \det(\mathbf{\Omega})^m \, \det(\mathbf{\Sigma})^m} \tag{47}$$

where ${}_0\tilde{F}_0\left(\cdot; \cdot; \cdot\right)$ is a complex hypergeometric function of three matrix arguments. To evaluate the integral in (46) we first expand ${}_0\tilde{F}_0(\cdot)$ in complex zonal polynomials [16] as

$$_0\tilde{F}_0\left(-\mathbf{\Omega}^{-1}, \mathbf{\Sigma}^{-1}, \mathbf{\Lambda}\right) = \sum_{k=0}^{\infty} \sum_{\mathcal{K}} \frac{\tilde{C}_{\mathcal{K}}(-\mathbf{\Omega}^{-1})\tilde{C}_{\mathcal{K}}(\mathbf{\Sigma}^{-1})\tilde{C}_{\mathcal{K}}(\mathbf{\Lambda})}{k! \tilde{C}_{\mathcal{K}}(\mathbf{I}_m)^2} \tag{48}$$

where the inner sum is over all partitions $\mathcal{K} = (k_1, \ldots, k_m)$ with $k_1 \geq k_2 \geq \ldots k_m \geq 0$, and $k_1 + \ldots + k_m = k$. Using the character representation for complex zonal polynomials [16]

$$\tilde{C}_{\mathcal{K}}(\mathbf{\Lambda}) = \chi_{[\mathcal{K}]}(1)\chi_{\{\mathcal{K}\}}(\mathbf{\Lambda}) \tag{49}$$

and

$$\tilde{C}_{\mathcal{K}}(\mathbf{I}_m) = \frac{\chi_{[\mathcal{K}]}(1)^2 \, \Gamma_m(m, \mathcal{K})}{k! \, \Gamma_m(m)} \tag{50}$$

where $\chi_{\{\mathcal{K}\}}(\cdot)$ is the character of the representation $\{\mathcal{K}\}$ of the linear group, $\chi_{[\mathcal{K}]}(1)$ is the dimension of the representation $[\mathcal{K}]$ of the symmetric group, and

$$\Gamma_m(m, \mathcal{K}) = \prod_{i=1}^{m} \Gamma(m + k_i - i + 1) \tag{51}$$

we can write

$$
{}_0\tilde{F}_0\left(-\mathbf{\Omega}^{-1},\mathbf{\Sigma}^{-1},\mathbf{\Lambda}\right) = \sum_{k=0}^{\infty}\sum_{\mathcal{K}}\frac{k!\,\Gamma_m(m)^2}{\Gamma_m(m,\mathcal{K})^2}\frac{\chi_{\{\mathcal{K}\}}(-\mathbf{\Omega}^{-1})\chi_{\{\mathcal{K}\}}(\mathbf{\Sigma}^{-1})\chi_{\{\mathcal{K}\}}(\mathbf{\Lambda})}{\chi_{[\mathcal{K}]}(1)}\,. \tag{52}
$$

Now we apply Weyl's formulas [25]

$$
\chi_{[\mathcal{K}]}(1) = k!\frac{\prod_{i<j}^m (k_i - k_j - i + j)}{\Gamma_m(m,\mathcal{K})} \tag{53}
$$

and[3]

$$
\chi_{\{\mathcal{K}\}}(\mathbf{\Lambda}) = \frac{\det\left(\lambda_i^{k_j+m-j}\right)}{\det\left(\lambda_i^{m-j}\right)} = (-1)^{m(m-1)/2}\frac{\det\left(\lambda_i^{k_j+m-j}\right)}{\Delta_m(\mathbf{\Lambda})} \tag{54}
$$

to give

$$
{}_0\tilde{F}_0\left(-\mathbf{\Omega}^{-1},\mathbf{\Sigma}^{-1},\mathbf{\Lambda}\right) = \frac{\Gamma_m(m)^2}{\prod_{i<j}^m\left[\left(\frac{1}{\omega_j}-\frac{1}{\omega_i}\right)\left(\frac{1}{\sigma_i}-\frac{1}{\sigma_j}\right)\right]}
$$
$$
\times\sum_{k=0}^{\infty}\sum_{\mathcal{K}}\frac{\det\left(\left(-\frac{1}{\omega_i}\right)^{k_j+m-j}\right)\det\left(\left(\frac{1}{\sigma_i}\right)^{k_j+m-j}\right)\det\left(\lambda_i^{k_j+m-j}\right)}{\Gamma\left(m,\mathcal{K}\right)\prod_{i<j}^m\left(k_j-k_i-j+i\right)\Delta_m(\mathbf{\Lambda})}\,. \tag{55}
$$

Now changing from $\mathcal{K}$ to *strictly ordered* partitions $\mathcal{K}_o = (\tilde{k}_1,\ldots,\tilde{k}_m)$ with $\tilde{k}_1 > \ldots > \tilde{k}_m \geq 0$ (i.e. such that $k_i + m - i \to \tilde{k}_i$), and using

$$
\prod_{i<j}^m\left(\frac{1}{\omega_j}-\frac{1}{\omega_i}\right) = \frac{\prod_{i<j}^m\left(\omega_i-\omega_j\right)}{\prod_{i=1}^m\omega_i^{m-1}} \tag{56}
$$

we obtain

$$
{}_0\tilde{F}_0\left(-\mathbf{\Omega}^{-1},\mathbf{\Sigma}^{-1},\mathbf{\Lambda}\right) = \frac{(-1)^{m(m-1)/2}\Gamma_m(m)^2\det\left(\mathbf{\Omega}\right)^{m-1}\det\left(\mathbf{\Sigma}\right)^{m-1}}{\Delta_m\left(\mathbf{\Omega}\right)\Delta_m\left(\mathbf{\Sigma}\right)}
$$
$$
\times\sum_{k=0}^{\infty}\sum_{\mathcal{K}_o}\frac{\det\left(\left(-\frac{1}{\omega_i}\right)^{\tilde{k}_j}\right)\det\left(\left(\frac{1}{\sigma_i}\right)^{\tilde{k}_j}\right)\det\left(\lambda_i^{\tilde{k}_j}\right)}{\left(\prod_{i=1}^m\tilde{k}_i!\right)\Delta_m(\mathcal{K}_o)\Delta_m(\mathbf{\Lambda})}\,. \tag{57}
$$

Substituting (57) into (47) and simplifying, the joint eigenvalue p.d.f. becomes

$$
f(\mathbf{\Lambda}) = \frac{(-1)^{m(m-1)/2}}{\det(\mathbf{\Omega})\det(\mathbf{\Sigma})\Delta_m(\mathbf{\Omega})\Delta_m(\mathbf{\Sigma})}
$$
$$
\times\sum_{k=0}^{\infty}\sum_{\mathcal{K}_o}\frac{\det\left(\left(-\frac{1}{\omega_i}\right)^{\tilde{k}_j}\right)\det\left(\left(\frac{1}{\sigma_i}\right)^{\tilde{k}_j}\right)\det\left(\lambda_i^{\tilde{k}_j}\right)\Delta_m(\mathbf{\Lambda})}{\left(\prod_{i=1}^m\tilde{k}_i!\right)\Delta_m(\mathcal{K}_o)}\,. \tag{58}
$$

---

[3]Here we introduce the compact notation for the determinant of a matrix, written in terms of the $(i,j)^{\text{th}}$ element.

Now substituting (58) into (46), the c.d.f. of $\lambda_m$ can be written as

$$F_{\lambda_m}(x) = \frac{(-1)^{m(m-1)/2}}{\det(\mathbf{\Omega})\det(\mathbf{\Sigma})\Delta_m(\mathbf{\Omega})\Delta_m(\mathbf{\Sigma})} \sum_{k=0}^{\infty} \sum_{\mathcal{K}_o} \frac{\det\left(\left(-\frac{1}{\omega_i}\right)^{\tilde{k}_j}\right)\det\left(\left(\frac{1}{\sigma_i}\right)^{\tilde{k}_j}\right)\mathcal{J}}{\left(\prod_{i=1}^{m}\tilde{k}_i!\right)\Delta_m(\mathcal{K}_o)} \quad (59)$$

where

$$\mathcal{J} = \int_{\mathcal{D}} \det\left(\lambda_i^{\tilde{k}_j}\right)\Delta_m(\mathbf{\Lambda})\,d\mathbf{\Lambda}\;. \quad (60)$$

To evaluate this integral, we note that $\Delta_m(\mathbf{\Lambda}) = \det\left(\lambda_i^{j-1}\right)$, and apply [26, Corr. 2] to obtain

$$\mathcal{J} = \det\left(\int_0^x t^{\tilde{k}_j+i-1}dt\right)$$

$$= x^{\frac{m(m+1)}{2}+\tilde{k}}\det\left(\frac{1}{\tilde{k}_j+i}\right)\;. \quad (61)$$

Now, in order to remove the infinite sum over partitions in (59), we must manipulate the right-hand determinant in (61) as follows

$$\det\left(\frac{1}{\tilde{k}_j+i}\right) = \det\begin{pmatrix} \frac{1}{k_1+1} & \frac{1}{k_2+1} & \cdots & \frac{1}{k_m+1} \\ \frac{1}{k_1+2} & \frac{1}{k_2+2} & \cdots & \frac{1}{k_m+2} \\ \vdots & \vdots & \ddots & \vdots \\ \frac{1}{k_1+m} & \frac{1}{k_2+m} & \cdots & \frac{1}{k_m+m} \end{pmatrix} \quad (62)$$

$$= \det\begin{pmatrix} \frac{1}{k_1+1} & \frac{\tilde{k}_1-\tilde{k}_2}{(k_1+1)(k_2+1)} & \cdots & \frac{\tilde{k}_1-\tilde{k}_m}{(k_1+1)(k_m+1)} \\ \frac{1}{k_1+2} & \frac{\tilde{k}_1-\tilde{k}_2}{(k_1+2)(k_2+2)} & \cdots & \frac{\tilde{k}_1-\tilde{k}_m}{(k_1+2)(k_m+2)} \\ \vdots & \vdots & \ddots & \vdots \\ \frac{1}{k_1+m} & \frac{\tilde{k}_1-\tilde{k}_2}{(k_1+m)(k_2+m)} & \cdots & \frac{\tilde{k}_1-\tilde{k}_m}{(k_1+m)(k_m+m)} \end{pmatrix} \quad (63)$$

$$= \prod_{i=2}^{m}(\tilde{k}_1-\tilde{k}_i)\prod_{i=1}^{m}\left(\frac{1}{\tilde{k}_1+i}\right)\det\begin{pmatrix} 1 & \frac{1}{k_2+1} & \cdots & \frac{1}{k_m+1} \\ 1 & \frac{1}{k_2+2} & \cdots & \frac{1}{k_m+2} \\ \vdots & \vdots & \ddots & \vdots \\ 1 & \frac{1}{k_2+m} & \cdots & \frac{1}{k_m+m} \end{pmatrix} \quad (64)$$

$$= \prod_{i=2}^{m}(\tilde{k}_1-\tilde{k}_i)\prod_{i=1}^{m}\left(\frac{1}{\tilde{k}_1+i}\right)\det\begin{pmatrix} 1 & \frac{1}{k_2+1} & \cdots & \frac{1}{k_m+1} \\ 0 & \frac{-1}{(k_2+1)(k_2+2)} & \cdots & \frac{-1}{(k_m+1)(k_m+2)} \\ 0 & \frac{-2}{(k_2+1)(k_2+3)} & \cdots & \frac{-2}{(k_m+1)(k_m+3)} \\ \vdots & \vdots & \ddots & \vdots \\ 0 & \frac{-(m-1)}{(k_2+1)(k_2+m)} & \cdots & \frac{-(m-1)}{(k_m+1)(k_m+m)} \end{pmatrix} \quad (65)$$

$$= \Gamma(m) \prod_{i=2}^{m} \left( \frac{\tilde{k}_i - \tilde{k}_1}{\tilde{k}_i + 1} \right) \prod_{i=1}^{m} \left( \frac{1}{\tilde{k}_1 + i} \right) \det \begin{pmatrix} \frac{1}{\tilde{k}_2 + 2} & \cdots & \frac{1}{\tilde{k}_m + 2} \\ \frac{1}{\tilde{k}_2 + 3} & \cdots & \frac{1}{\tilde{k}_m + 3} \\ \vdots & \ddots & \vdots \\ \frac{1}{\tilde{k}_2 + m} & \cdots & \frac{1}{\tilde{k}_m + m} \end{pmatrix} \tag{66}$$

Note that in (63) and (64) we have subtracted the first column from all other columns and removed common factors. In (65) and (66) we have then subtracted the first row in the determinant from all other rows and removed common factors. Notice that the determinant in (66) is a principle submatrix of the determinant in (62). Applying the same sequence of operations ($m - 1$ times) to the determinant in (66), we obtain

$$\det \left( \frac{1}{\tilde{k}_j + i} \right) = \Gamma_m(m) \, \Delta_m(\mathcal{K}_o) \prod_{i,j=1}^{m} \left( \frac{1}{\tilde{k}_i + j} \right) . \tag{67}$$

Substituting (67) into (61) and (59) and simplifying gives

$$F_{\lambda_m}(x) = \frac{(-1)^{m(m-1)/2} \, \Gamma_m(m) x^{m(m+1)/2}}{\det(\boldsymbol{\Omega}) \det(\boldsymbol{\Sigma}) \, \Delta_m(\boldsymbol{\Omega}) \, \Delta_m(\boldsymbol{\Sigma})} \sum_{k=0}^{\infty} \sum_{\mathcal{K}_o} \det \left( \left( -\frac{1}{\omega_i} \right)^{\tilde{k}_j} \right) \det \left( \left( \frac{1}{\sigma_i} \right)^{\tilde{k}_j} \right) \prod_{i=1}^{m} g(\tilde{k}_i) \tag{68}$$

where

$$g(\tilde{k}_i) = \frac{x^{\tilde{k}_i}}{\tilde{k}_i!} \prod_{j=1}^{m} \left( \frac{1}{\tilde{k}_i + j} \right) . \tag{69}$$

Now we apply the Cauchy-Binet formula [27] to give

$$F_{\lambda_m}(x) = \frac{(-1)^{m(m-1)/2} \, \Gamma_m(m) \, x^{m(m+1)/2}}{\det(\boldsymbol{\Omega}) \det(\boldsymbol{\Sigma}) \, \Delta_m(\boldsymbol{\Omega}) \, \Delta_m(\boldsymbol{\Sigma})} \det \left( \sum_{k=0}^{\infty} \left( -\frac{1}{\omega_i \sigma_j} \right)^{k} g(k) \right) \tag{70}$$

and deal with the infinite sum as follows

$$\begin{aligned}
\sum_{k=0}^{\infty} \left( -\frac{1}{\omega_i \sigma_j} \right)^{k} g(k) &= \sum_{k=0}^{\infty} \frac{\left( -\frac{x}{\omega_i \sigma_j} \right)^{k}}{k!} \prod_{j=1}^{m} \left( \frac{1}{k + j} \right) \\
&= \sum_{k=0}^{\infty} \frac{\left( -\frac{x}{\omega_i \sigma_j} \right)^{k}}{k!} \frac{\Gamma(k+1)}{\Gamma(m+k+1)} \\
&= \frac{1}{m!} {}_1F_1 \left( 1; m+1; -\frac{x}{\omega_i \sigma_j} \right)
\end{aligned} \tag{71}$$

where $_1F_1(\cdot)$ is the confluent hypergeometric function. Applying the identity (see [22, Eq. 6.5.2] and [22, Eq. 6.5.12])

$$_1F_1(1; b; z) = \Gamma(b) z^{1-b} e^z \, \mathcal{P}(b-1; z) \tag{72}$$

in (71), and then substituting into (70) yields

$$F_{\lambda_m}(x) = \frac{(-1)^m \, \Gamma_m(m) \det(\boldsymbol{\Omega})^{m-1} \det(\boldsymbol{\Sigma})^{m-1}}{\Delta_m(\boldsymbol{\Omega}) \, \Delta_m(\boldsymbol{\Sigma}) \, (-x)^{m(m-1)/2}} \det\left(e^{-\frac{x}{\omega_i \sigma_j}} \mathcal{P}\left(m; -\frac{x}{\omega_i \sigma_j}\right)\right) \,. \tag{73}$$

This establishes the result for square matrices.

To obtain the result for rectangular matrices (i.e. for $n < m$, and $n \times n$ matrix $\boldsymbol{\Omega}$), we follow an approach of [15], and consider an auxiliary (expanded) system with $m \times m$ matrix $\tilde{\boldsymbol{\Omega}}$ with eigenvalues $(\tilde{\omega}_1, \ldots, \tilde{\omega}_m) = (\epsilon_1, \ldots, \epsilon_\tau, \omega_1, \ldots, \omega_n)$, for which (73) holds, and take limits of (73) as $\epsilon_1 \to 0, \ldots, \epsilon_\tau \to 0$. We then have

$$F_{\lambda_m}(x) = \frac{(-1)^m \, \Gamma_m(m) \det(\boldsymbol{\Sigma})^{m-1}}{\Delta_m(\boldsymbol{\Sigma})(-x)^{m(m-1)/2}} \mathcal{L} \tag{74}$$

where

$$\mathcal{L} = \lim_{\tilde{\omega}_1 \to 0, \ldots, \tilde{\omega}_\tau \to 0} \frac{\det\left(\tilde{\boldsymbol{\Omega}}\right)^{m-1}}{\Delta_m\left(\tilde{\boldsymbol{\Omega}}\right)} \det\left(e^{-\frac{x}{\tilde{\omega}_i \sigma_j}} \mathcal{P}\left(m; -\frac{x}{\tilde{\omega}_i \sigma_j}\right)\right) \,. \tag{75}$$

To take these limits, we start by noting that

$$e^{-\frac{x}{\tilde{\omega}_i \sigma_j}} \mathcal{P}\left(m; -\frac{x}{\tilde{\omega}_i \sigma_j}\right) = e^{-\frac{x}{\tilde{\omega}_i \sigma_j}} \left(1 - \frac{\Gamma\left(m, -\frac{x}{\tilde{\omega}_i \sigma_j}\right)}{\Gamma(m)}\right) \tag{76}$$

where $\Gamma(\cdot, \cdot)$ is the upper incomplete gamma function, and use [22, Eq. 6.5.32] to obtain

$$\lim_{\omega_i \to 0} e^{-\frac{x}{\tilde{\omega}_i \sigma_j}} \mathcal{P}\left(m; -\frac{x}{\tilde{\omega}_i \sigma_j}\right) = \frac{-1}{\Gamma(m)} \left(-\frac{x}{\tilde{\omega}_i \sigma_j}\right)^{m-1} \left(1 + \frac{m-1}{-\frac{x}{\tilde{\omega}_i \sigma_j}} + \frac{(m-1)(m-2)}{(-\frac{x}{\tilde{\omega}_i \sigma_j})^2} + \ldots\right) \,. \tag{77}$$

We then consider each of the $\tau$ limits in turn, starting with $\tilde{\omega}_1 \to 0$. We replace the first row in the right-hand determinant in (75) with the first term in the expansion in (77) (i.e. $-\frac{1}{\Gamma(m)}\left(-\frac{x}{\tilde{\omega}_1 \sigma_j}\right)^{m-1}$). For the denominator we have

$$\lim_{\tilde{\omega}_1 \to 0} \Delta_m(\tilde{\boldsymbol{\Omega}}) = \Delta_{m-1}(\tilde{\omega}_2, \ldots, \tilde{\omega}_m) \prod_{j=2}^m \tilde{\omega}_j \,. \tag{78}$$

Now, factoring the numerator determinant yields a finite ratio for $\mathcal{L}$ in (75) in the limit $\tilde{\omega}_1 \to 0$,

given by

$$\mathcal{L} = \frac{(-1)^m \, x^{m-1}}{\Gamma(m)} \lim_{\tilde{\omega}_2 \to 0, \dots, \tilde{\omega}_\tau \to 0} \frac{\left(\prod_{j=2}^m \tilde{\omega}_j\right)^{m-2} \det\left(\boldsymbol{\Xi}_1(x)\right)}{\Delta_{m-1}(\tilde{\omega}_2, \dots, \tilde{\omega}_m)} \tag{79}$$

where $\boldsymbol{\Xi}_\ell(x)$ is an $m \times m$ matrix with $(i,j)^{\text{th}}$ element

$$\left(\boldsymbol{\Xi}_\ell(x)\right)_{i,j} = \begin{cases} \left(\frac{1}{\sigma_j}\right)^{m-i} & , \, i \le \ell \\ e^{-\frac{x}{\tilde{\omega}_i \sigma_j}} \mathcal{P}\left(m; -\frac{x}{\tilde{\omega}_i \sigma_j}\right) & , \, i > \ell \end{cases} \tag{80}$$

Now we take the limit as $\tilde{\omega}_2 \to 0$. In this case the leading divergence in the numerator comes from the second term in the expansion (77) (i.e. taking the first term results in a determinant of zero). Hence, we can replace the second row of $\boldsymbol{\Xi}_1(x)$ with $\frac{(-1)^{m-1}}{\Gamma(m-1)} \left(\frac{x}{\tilde{\omega}_i \sigma_j}\right)^{m-2}$. For the denominator we have

$$\lim_{\tilde{\omega}_2 \to 0} \Delta_{m-1}(\tilde{\boldsymbol{\Omega}}) = \Delta_{m-2}(\tilde{\omega}_3, \dots, \tilde{\omega}_m) \prod_{j=3}^m \tilde{\omega}_j \; . \tag{81}$$

Hence, factoring the numerator determinant again yields a finite ratio for $\mathcal{L}$ as $\tilde{\omega}_2 \to 0$, given by

$$\mathcal{L} = \frac{(-1)^{m+(m-1)} \, x^{(m-1)+(m-2)}}{\Gamma(m)\Gamma(m-1)} \lim_{\tilde{\omega}_3 \to 0, \dots, \tilde{\omega}_\tau \to 0} \frac{\left(\prod_{j=3}^m \tilde{\omega}_j\right)^{m-3} \det\left(\boldsymbol{\Xi}_2(x)\right)}{\Delta_{m-2}(\tilde{\omega}_3, \dots, \tilde{\omega}_m)} \; . \tag{82}$$

Continuing this procedure until all limits have been calculated, and simplifying the result, yields

$$\mathcal{L} = \frac{(-1)^{\frac{n(n+1)}{2}} \, x^{-\frac{n(n-1)}{2}} \det\left(\boldsymbol{\Omega}\right)^{n-1} \det\left(\boldsymbol{\Xi}_\tau(x)\right)}{\Gamma(m) \dots \Gamma(m-\tau+1) \, \Delta_n(\boldsymbol{\Omega})} \; . \tag{83}$$

Now noting that $\boldsymbol{\Xi}_\tau(x) = \boldsymbol{\Psi}(x)$, where $\boldsymbol{\Psi}(x)$ is defined in the theorem, and substituting (83) into (74), we perform some basic manipulations to obtain the desired result in (1). □

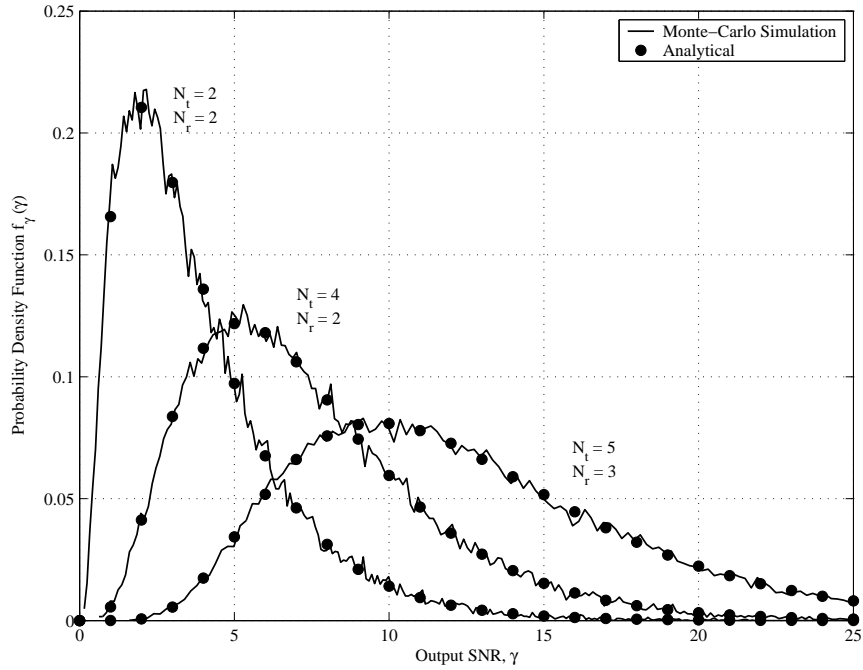

Fig. 1. P.d.f. of the output SNR of MIMO-MRC in double-correlated Rayleigh channels, for $\bar{\gamma} = 0$ dB. Correlation parameters are $\theta_r = \theta_t = \frac{\pi}{2}$, $d = \frac{1}{2}$, $\sigma_r^2 = \frac{\pi}{64}$, and $\sigma_t^2 = \frac{\pi}{16}$.

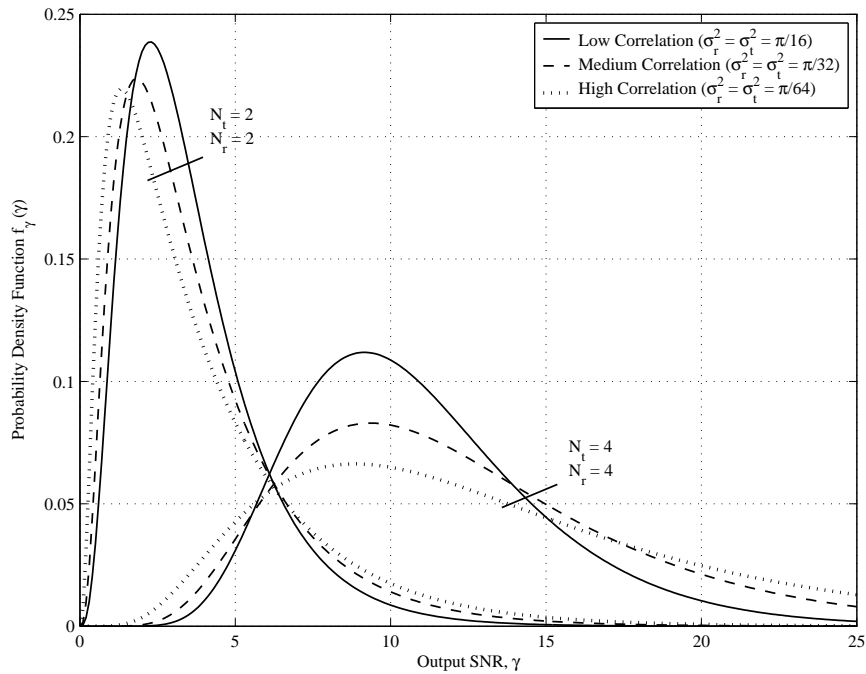

Fig. 2. P.d.f. of the output SNR of MIMO-MRC in various double-correlated Rayleigh channels, for $\bar{\gamma} = 0$ dB.

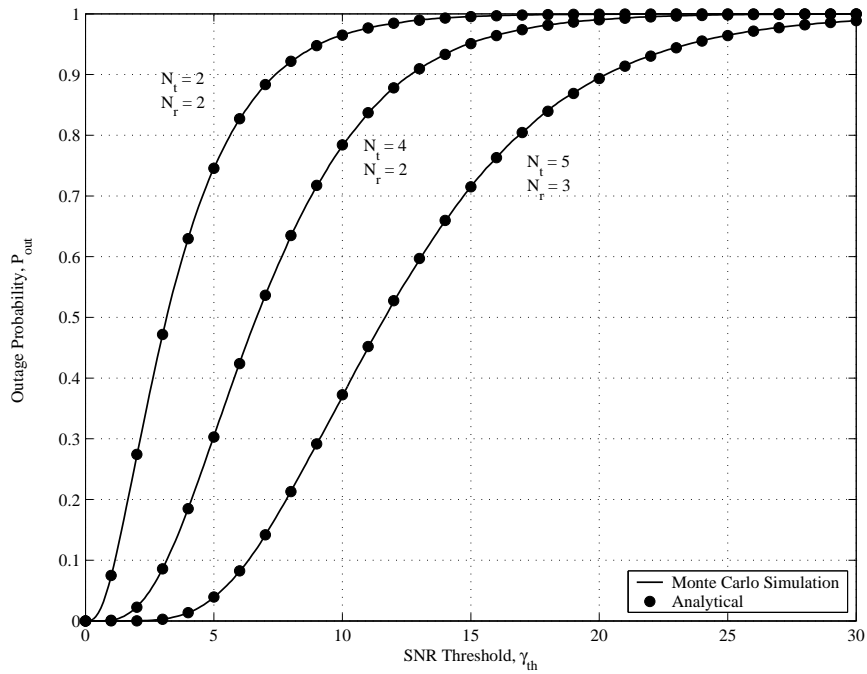

Fig. 3.   Outage probability of MIMO-MRC in double-correlated Rayleigh channels, for $\bar{\gamma} = 0$ dB. Correlation parameters are $\theta_r = \theta_t = \frac{\pi}{2}$, $d = \frac{1}{2}$, $\sigma_r^2 = \frac{\pi}{64}$, and $\sigma_t^2 = \frac{\pi}{16}$.

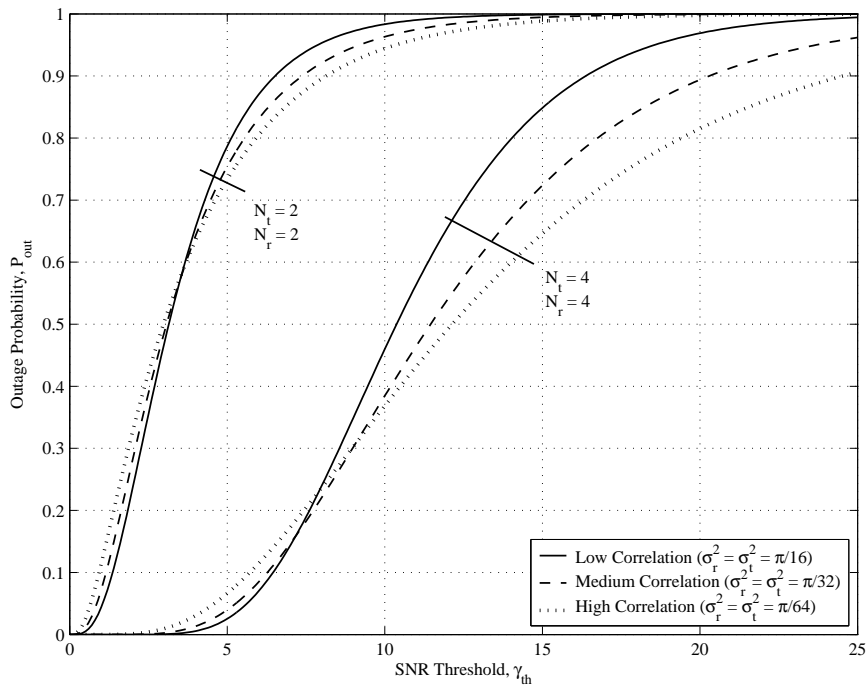

Fig. 4.   Outage probability of MIMO-MRC in various double-correlated Rayleigh channels, for $\bar{\gamma} = 0$ dB.

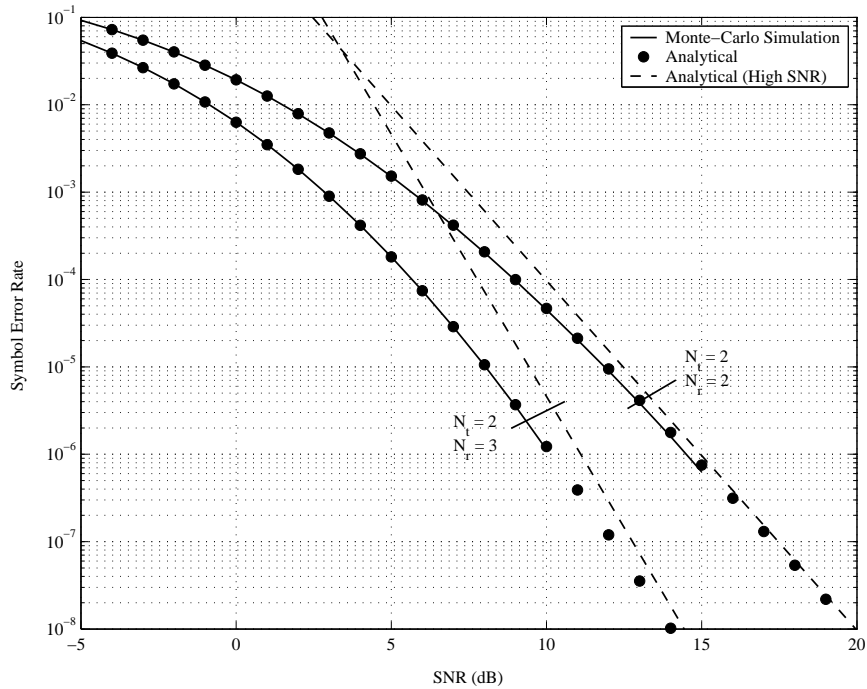

Fig. 5.   Symbol Error Rate of MIMO-MRC in double-correlated Rayleigh channels with BPSK modulation. Correlation parameters are $\theta_r = \theta_t = \frac{\pi}{2}$, $d = \frac{1}{2}$, $\sigma_r^2 = \frac{\pi}{16}$, and $\sigma_t^2 = \frac{\pi}{32}$.

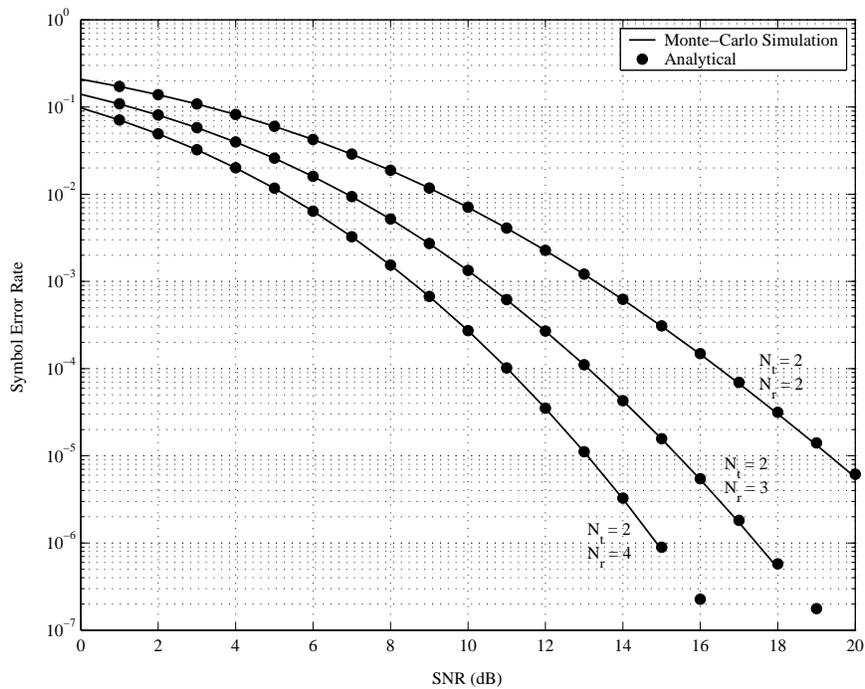

Fig. 6.   Symbol Error Rate of MIMO-MRC in double-correlated Rayleigh channels with 4-PAM modulation. Correlation parameters are $\theta_r = \theta_t = \frac{\pi}{2}$, $d = \frac{1}{2}$, $\sigma_r^2 = \frac{\pi}{16}$, and $\sigma_t^2 = \frac{\pi}{32}$.

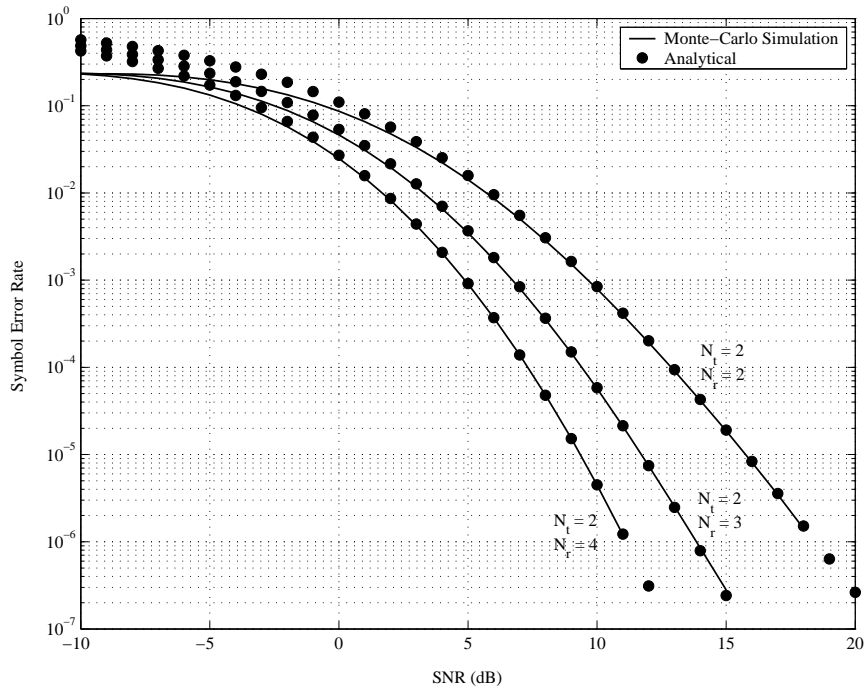

Fig. 7. Symbol Error Rate of MIMO-MRC in double-correlated Rayleigh channels with QPSK modulation. Correlation parameters are $\theta_r = \theta_t = \frac{\pi}{2}$, $d = \frac{1}{2}$, $\sigma_r^2 = \frac{\pi}{16}$, and $\sigma_t^2 = \frac{\pi}{32}$.

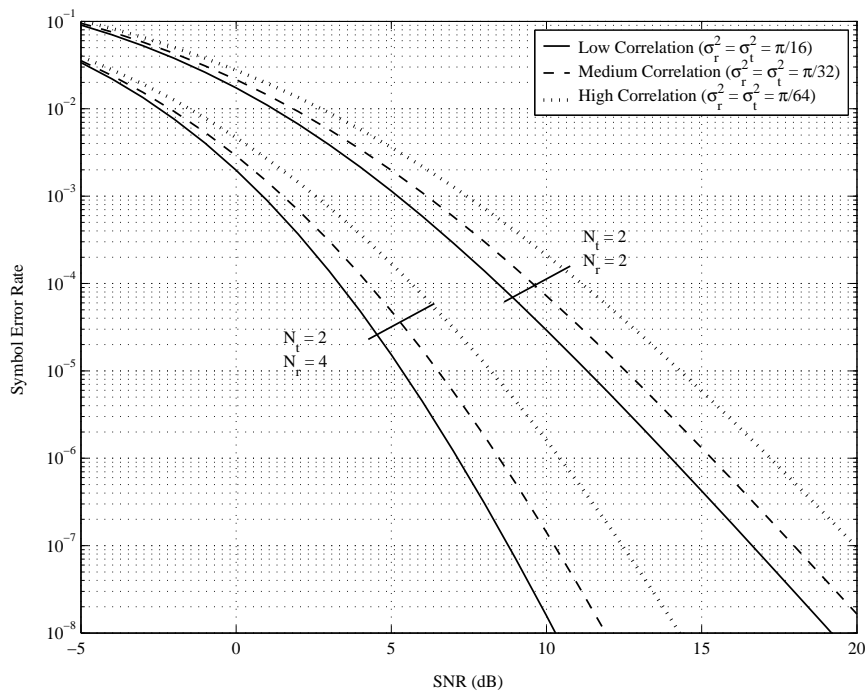

Fig. 8. Symbol Error Rate of MIMO-MRC in various double-correlated Rayleigh channels with BPSK modulation.